LUXEMBOURG INSTITUTE OF SCIENCE AND TECHNOLOGY

# AN INTEGRATED CONCEPTUAL MODEL FOR INFORMATION SYSTEM SECURITY RISK MANAGEMENT AND ENTERPRISE ARCHITECTURE MANAGEMENT BASED ON TOGAF, ARCHIMATE, IAF AND DODAF

TECHNICAL REPORT

**Luxembourg Institute of Science and Technology (LIST)**
IT for Innovative Services
5, avenue des HautsFourneaux | L-4362 Esch/Alzette
Tel: (+352) 275 888 – 1 | LIST.lu

# About LIST

The Luxembourg Institute of Science and Technology (LIST) is a mission-driven Research and Technology Organisation (RTO) that develops advanced technologies and delivers innovative products and services to industry and society.

As a major engine of the diversification and growth of Luxembourg's economy through innovation, LIST supports the deployment of a number of solutions to a wide range of sectors, including energy, IT, telecommunications, environment, agriculture, and advanced manufacturing at national and European level.

Thanks to its location in an exceptional collaborative environment, namely the Belval Innovation Campus, LIST accelerates time to market by maximizing synergies with different actors, including the university, the national funding agency and industrial clusters.

# Acknowledgments

This work has been supported by the National Research Fund, Luxembourg, and financed by the ENTRI project (C14/IS/8329158).

# Contact details

For more information about this document, please contact:

**Dr. Nicolas Mayer**

nicolas.mayer@list.lu

# Legal notice

Notice must be taken that this publication represents the views and interpretation of the authors, unless stated otherwise. This publication does not necessarily represent state-of-the-art and LIST may update it from time to time.

Third-party sources are quoted as appropriate. LIST is not responsible for the content of the external sources including external websites referenced in this publication.

This publication is intended for information purposes only. It must be accessible free of charge. Neither LIST nor any person acting on its behalf is responsible for the use that might be made of the information contained in this publication.

# Table of Contents

# Executive Summary

Risk management is today a major steering tool for any organization wanting to deal with Information System (IS) security. However, IS Security Risk Management (ISSRM) remains difficult to establish and maintain, mainly in a context of multi-regulations with complex and inter-connected IS. We claim that a connection with Enterprise Architecture Management (EAM) contributes to deal with these issues. A first step towards a better integration of both domains is to define an integrated EAM-ISSRM conceptual model [1], [2].

Among the steps of the research method followed to define such an integrated EAM-ISSRM conceptual, this technical report presents the whole outputs (through alignment tables) of the conceptual alignment between concepts used to model EA (based on ArchiMate, TOGAF, IAF and DoDAF) and concepts of the ISSRM domain model.

# Background

## The ISSRM Domain Model

In our preceding works, the concepts of ISSRM have been formalised as a domain model, i.e. a conceptual model depicting the studied domain [3]. The ISSRM domain model was designed from related literature [4]: risk management standards, security-related standards, security risk management standards and methods and security requirements engineering frameworks. The ISSRM domain model is composed of three groups of concepts: *Asset-related concepts*, *Risk-related concepts*, and *Risk treatment-related concepts*. Each of the concepts of the model has been defined and linked one to the other [4], as represented in Figure 1.

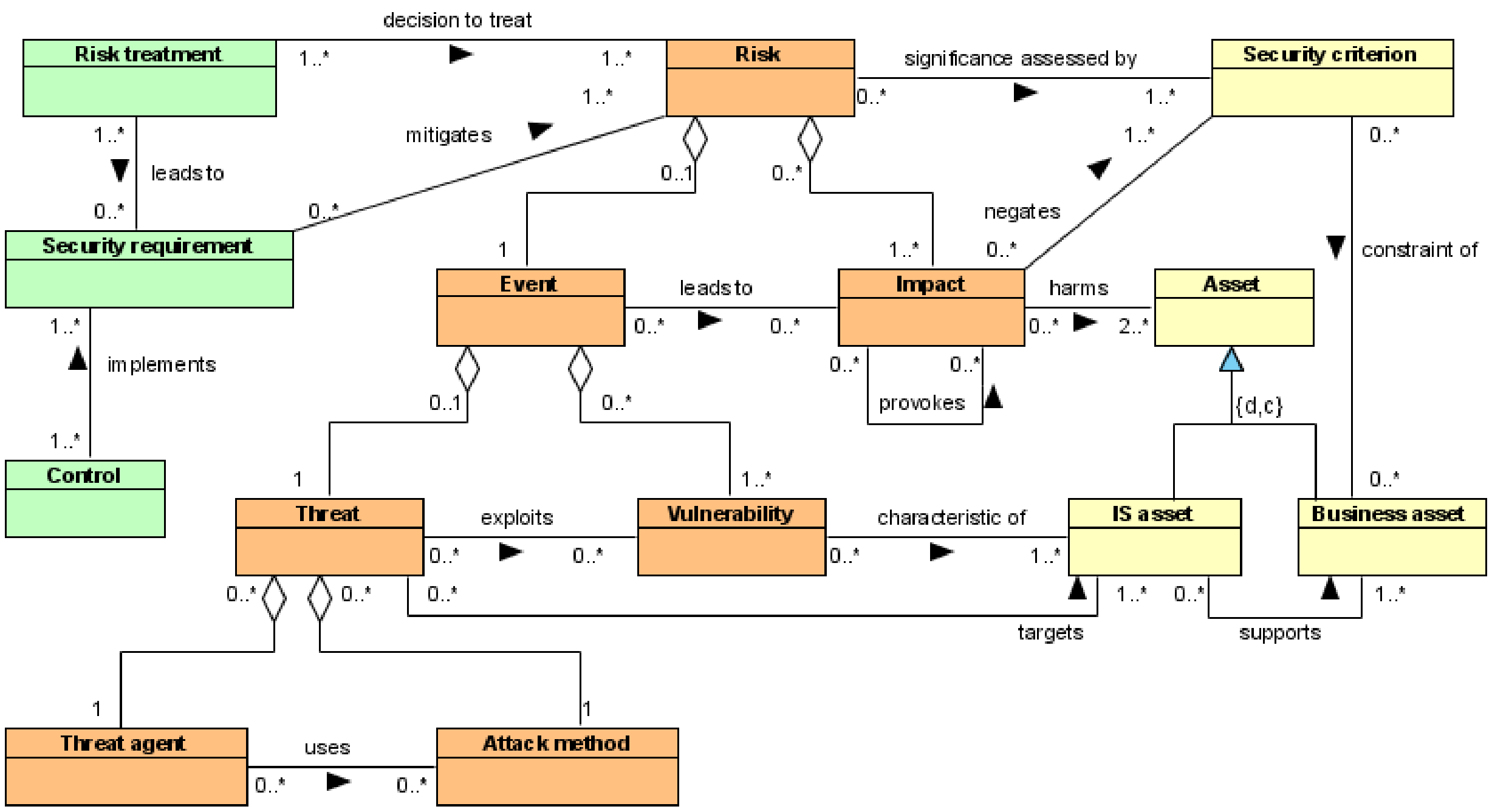

Figure 1: ISSRM domain model extracted from [4]

*Asset-related concepts* describe assets and the criteria which guarantee asset security. An *asset* is anything that has value to the organisation and is necessary for achieving its objectives. A *Business asset* describes information, processes, capabilities and skills inherent to the business and core mission of the organisation, having value for it. An *IS asset* is a component of the IS supporting Business assets like a database where information is stored. In our context, and as described in the ISSRM literature [4], an IS is a composition of hardware, software, network, people and facilities. A security criterion characterises a property or constraint on business assets describing their security needs, usually for confidentiality, integrity and availability.

*Risk-related concepts* present how the risk itself is defined. A *risk* is the combination of a *threat* with one or more *vulnerabilities* leading to a negative *impact* harming the assets. An *impact* describes the potential negative consequence of a risk that may harm assets of a system or organisation, when a *threat* (or the cause of a risk) is accomplished. An *event* is the combination of a threat and one or more vulnerabilities. A *vulnerability* describes a characteristic of an IS asset or group of IS assets that can constitute a weakness or a flaw in terms of IS security. A *threat* characterises a potential attack or incident, which targets one or more IS assets and may lead to the assets being harmed. A *threat agent* is an agent that can potentially cause harm to IS assets. An *attack method* is a standard means by which a threat agent carries out a threat.

*Risk treatment-related concepts* describe what decisions, requirements and controls should be defined and implemented in order to mitigate possible risks. A risk treatment is an intentional decision to treat identified risks. A security requirement is the refinement of a treatment decision to mitigate the risk. Controls (countermeasures or safeguards) are designed to improve security, specified by a security requirement, and implemented to comply with it.

## ArchiMate

To provide a uniform representation for diagrams that describe EA, the ArchiMate modelling language [5] has been developed by the Open Group, an industry consortium developing standards. It offers an integrated architectural approach to describe and visualize the different architecture domains and their underlying relations and dependencies. The role of the ArchiMate standard is to provide a graphical language for the representation of EA over time (i.e., including transformation and migration planning), as well as their motivation and rationale. The current version of the standard (used in this technical report) is 2.1 and its evolution is closely linked to the developments of the TOGAF standard [6] and the emerging results from The Open Group forums and work groups active in this area. It is today a widely accepted open standard for modelling EA [7], with a large user base and a variety of modelling tools that support it.

## TOGAF

TOGAF is a framework — a detailed method and a set of supporting tools — for developing an enterprise architecture [6]. It is a standard established and maintained by The Open Group, an industry consortium focused on IT standards. A key aspect of TOGAF is the TOGAF Architecture Development Method (ADM), a tested and repeatable process for developing architectures. The ADM includes establishing an architecture framework, developing architecture content, transitioning, and governing the realization of architectures. The TOGAF Architecture Content Framework (ACF) provides a structural model for architectural content, developed all along the different steps of the ADM, which allows major work products to be consistently defined, structured, and presented. The TOGAF ACF is structured according to its Content Metamodel. This metamodel is a single view that encompasses all four of the TOGAF architecture domains (Business, Data, Application; and Technology Architecture), and that defines a set of entities that allow architectural concepts to be captured, stored, filtered, queried, and represented in a way that supports consistency, completeness, and traceability. The TOGAF Content Metamodel and its associated glossary are of particular interest for the analysis performed in this paper. More information about TOGAF can be found in the TOGAF 9.1 reference book [6].

## DoDAF

The Department of Defense Architecture Framework (DoDAF), Version 2.02 [8] consists in a framework and conceptual model to develop architectures to facilitate the ability of Department of Defense (DoD) managers to make decisions.  It is an enterprise asset used to assess alignment with the missions of the DoD enterprise, to strengthen customer support, to support capability portfolio management (PfM), and to ensure that operational goals and strategies are met. DoDAF architecture development process provides guidance to the architect and development team. The process is data-centric rather than product or service-centric. The steps of the architecture development process are: Determine Intended Use of Architecture, Determine Scope of Architecture, Determine Data Required to Support Architecture Development, Collect, Organize, Correlate, and Store Architectural Data, Conduct Analyses in Support of Architecture Objectives, and Document Results in Accordance with Decision-Maker Needs.

The purpose of DoDAF is to define concepts and models usable in DoD core processes following three views:

The Conceptual Data Model defines the high-level data constructs from which Architectural Descriptions are created in non-technical terms, so that executives and managers at all levels can understand the data basis of Architectural Description.

The Logical Data Model adds technical information and clarifies relationships into an unambiguous usage definition.

The Physical Exchange Specification consists of the Logical Data Model with general data types specified and implementation attributes added.

## IAF

The Integrated Architecture Framework (IAF) is an enterprise architecture framework that covers business, information, information system and technology infrastructure [9]. This framework has been developed by Capgemini since the 1990s, from the experience of practicing architects on projects for clients across the group. The first version was released in 1996. It is a toolbox that contains processes, products, tools and techniques to create all types of architectures which are intended to shape businesses and the technology that supports it.

.

# Research Method

The research method followed to develop the EAM-ISSRM integrated model is composed of four steps detailed below. This technical report only focuses on the second step (i.e. the conceptual alignment between concepts used to model EA and concepts of the ISSRM domain model) by presenting the findings through alignment tables.

## Selection of relevant literature on EAM

The first step of the research method consists of selecting relevant literature on EAM that will be used to adapt and extend the ISSRM domain model with EA-related concepts. The literature on EAM is huge, and for our goal it is not necessary to perform a complete review of it. Indeed, to facilitate a high acceptance level of our extension by practitioners, we focus on conceptual models that are used in practice.

## Conceptual alignment between concepts used to model EA and concepts of the ISSRM domain model

The second step of the research method consists of identifying the semantic correspondence between concepts found in the selected literature on EAM and the concepts of the ISSRM domain model. This task is performed by a design group composed of experts of both domains, in order to consolidate as much as possible the alignment decisions. The approach followed is inspired by Zivkovic et al. [10]. Each relation between concepts is classified according to the following semantic mapping subtypes:

- **Equivalence:** concept A is semantically equivalent to concept B;
- **Generalisation**: concept A is a generalisation of concept B, i.e. concept B is a specific class of concept A;
- **Specialisation**: concept A is a specialisation of concept B, i.e. concept B is a generic class of concept A[1];
- **Aggregation**: concept A is composed of concept B, i.e. concept B is a part of concept A;
- **Composition**: concept A is composed of concept B (with strong ownership), i.e. concept B is a part of concept A and does only exist as part of concept A;
- **Association**: concept A is linked to concept B.

The output of this step is a table for each literature reference on EAM, highlighting the relations between its concepts and those of the ISSRM domain model, and illustrated, when applicable, with an example of use of the EA concepts in an ISSRM context. As a running example we use a model of a medical analysis laboratory, developed in a national project that aims to improve and facilitate RM in the medical sector.

## Design of the EAM-ISSRM integrated model

The third step of our research method consists of the design of an integrated EAM-ISSRM conceptual model. This integrated conceptual model is built incrementally, taking into account the different conceptual alignments performed for each studied literature reference. More specifically, we build a specific EAM-ISSRM integrated model for each studied literature reference in EAM and reconcile all of them afterwards.

## Validation of the EAM-ISSRM integrated model

In order to validate the result obtained, we get information about the utility and usability of the EAM-ISSRM integrated model by means of a focus group. This validation group is composed of experienced ISSRM practitioners who answer questions and perform exercises developed for assessing the utility and usability of the model. Members of the validation group are people not involved in the design stage of the EAM-ISSRM integrated conceptual model.

---

[1] Generalisation and Specialisation are opposite relations

# Conceptual Alignment between Concepts of EAM and Concepts of the ISSRM Domain Model

Four selected EAM methods have been investigated to define the EAM-ISSRM integrated model, namely ArchiMate 2.1 [5], TOGAF 9.1 [6], DoDAF [8], IAF [9]. Each conceptual alignment has been performed by a design group composed of five people. Three of them are ISSRM experts and two of them EAM experts. All of the members of the design group are researchers having a good theoretical knowledge of ISSRM and/or EAM. Moreover, two ISSRM experts are also experienced ISSRM practitioners (in total during the 10 last years, they have performed more than 20 real-world applications of ISSRM in organizations, ranging from SMEs to European institutions). Alignment decisions were taken only once a consensus has been found among the members of this design group.

## Presentation of the Running Example

In order to illustrate the conceptual alignment, a running example, excerpt from a complete modelling of the medical analysis laboratory sector, is presented for ArchiMate and TOGAF that are both suited to do so. This modelling has been developed in a national project aiming to improve and facilitate RM in the medical sector [11].

Desiring to build on a real-world example of modelling (and not a fictitious example built for this occasion), we are aware that these use-cases do not cover all of the studied concepts, however it allows to illustrate as much as possible the alignment achieved. It is also important to note that the concepts alignment was performed based on the definitions of the concepts and not only on the basis of the examples.

### ArchiMate

This excerpt, modelled with ArchiMate, details a specific activity of a medical analysis laboratory: the home blood sample collection. It is organised in four distinct views, namely a Business view (Figure 2) focussing on business part of “*Home Blood Analysis*”, a Motivation view (Figure 3) presenting the value proposition behind the development of “*Home Blood Analysis*”, an Information view (Figure 4) presenting information, and finally a Technology view (Figure 5) focussing on the technological architecture of a specific Business Function. These views have been developed to illustrate the majority of the constructs of the three ArchiMate layers, namely the Business Layer, Application Layer and Technology Layer, and represented with their standard visual aspect as depicted in ArchiMate 2.1 [5].

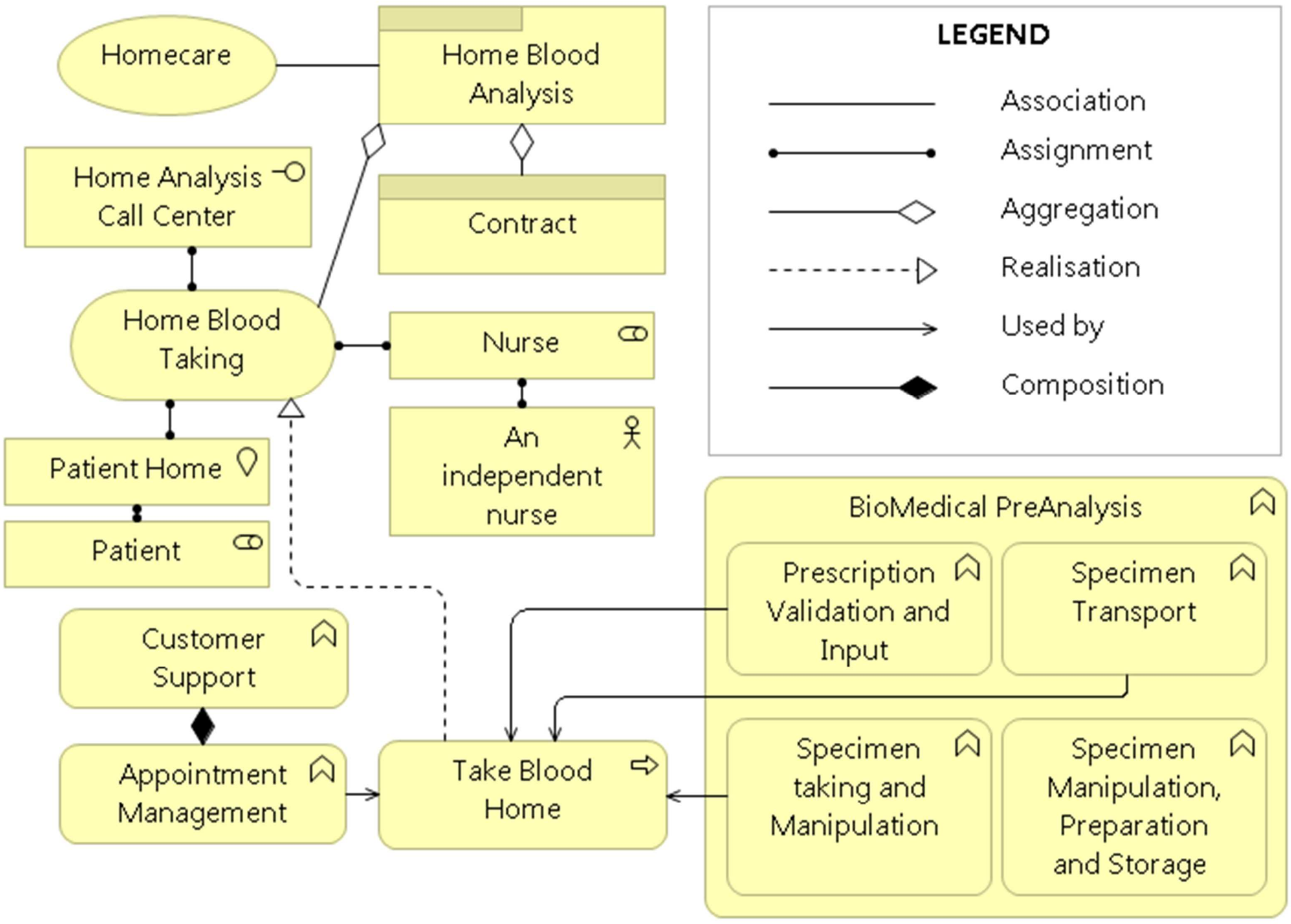

**Figure 2: Business view**

We are aware that this case does not cover all of the studied concepts, however the case is realistic, and it allows to illustrate as much as possible the alignment achieved. We have performed the concept alignment based on the definitions of the concepts from the specification of ArchiMate [5], and not only on the basis of this example. The example is used only to illustrate the use of ArchiMate constructs (which are very generic) in a context of ISSRM.

The Business view (Figure 2) is focused on "*Home Blood Analysis*" as a product proposed by the laboratory to its patients. This product is composed of services such as "*Home Blood taking*", themselves composed of business processes ("*Take Blood Home*") and business functions ("*Appointement Management*", "*BioMedical PreAnalysis*").

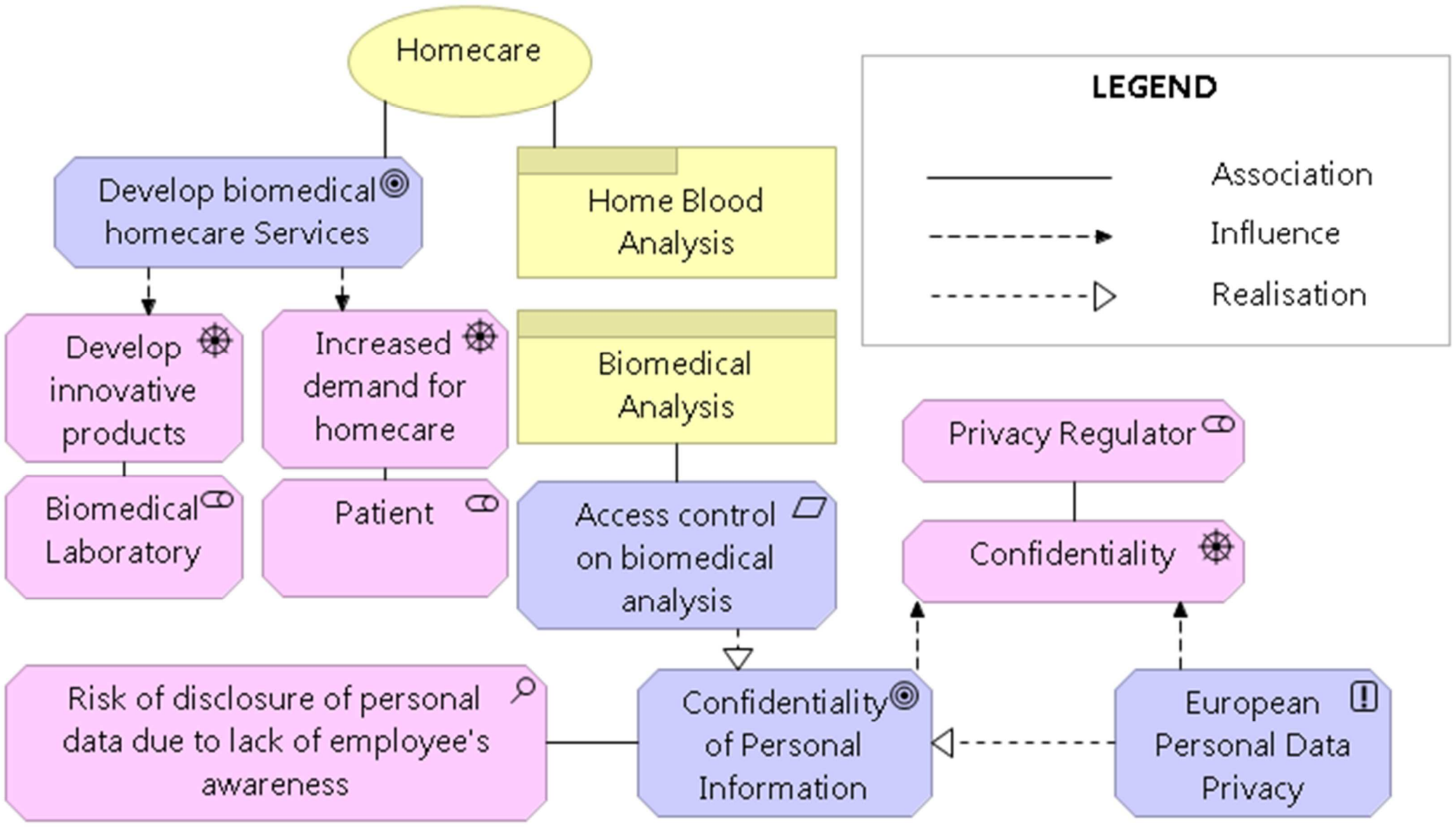

**Figure 3: Motivation view**

The Motivation view (Figure 3) introduces the value proposition of this product. To develop biomedical homecare services is considered as a goal for the laboratory. In this frame, it takes into account its motivations and business context, such as an increased demand of patients for homecare or the European regulation about personal data protection.

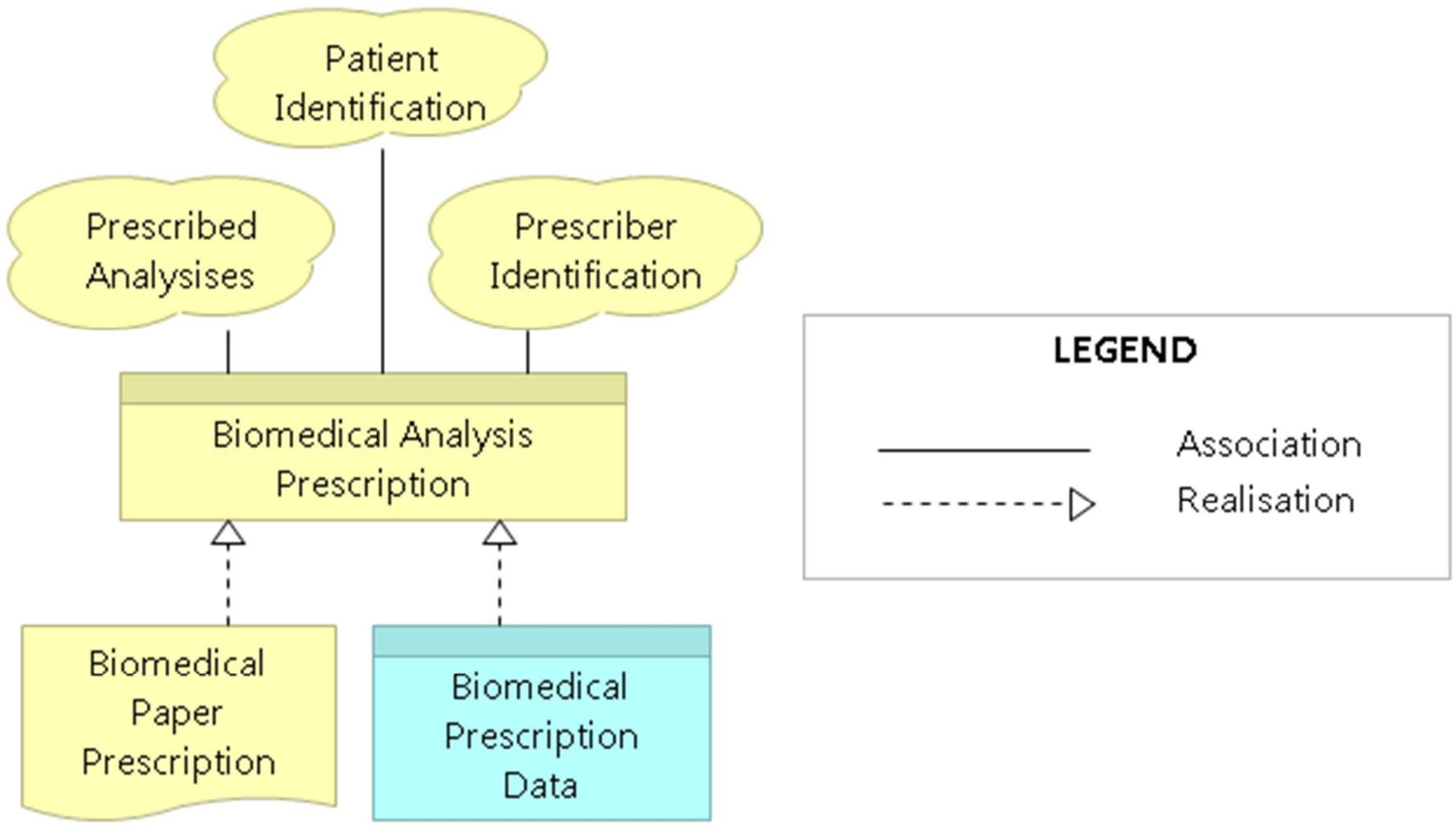

Figure 4: Semantic view

Then, the Information view (Figure 4) focuses on "*Biomedical Analysis Prescription*" that is a key informational asset of the laboratory. It makes clear which medical information is included in this kind of prescription and that its support can be digital or paper-based. Finally, the Technology view (Figure 5) models how "*Biomedical Analysis Prescriptions*" are established through software applications at the Application Layer, while the Technology Layer exhibits the underlying technical architecture.

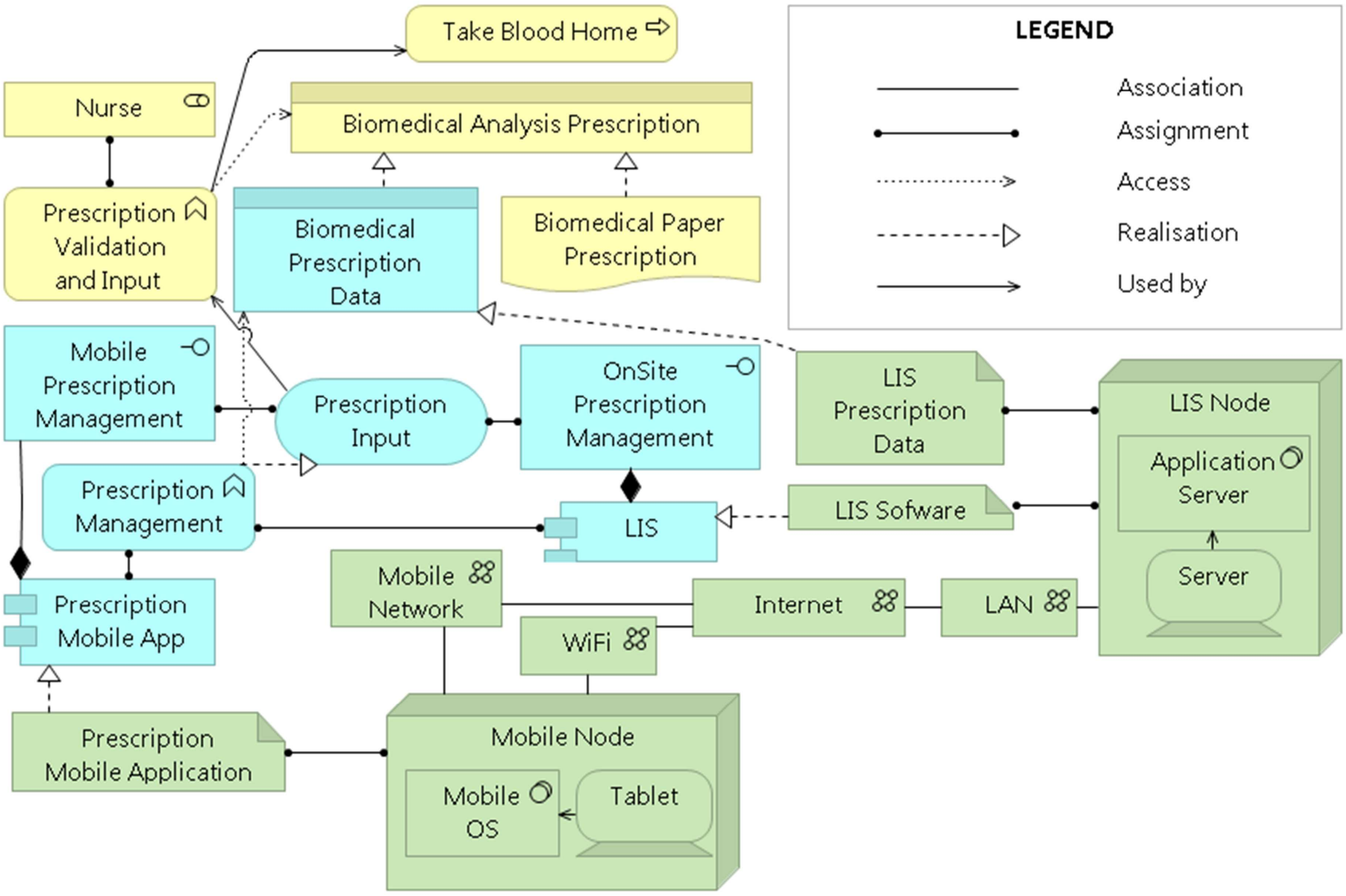

Figure 5: Technology view

## TOGAF

This excerpt, modelled with TOGAF 9, represents some of the components of medical analysis activities for hospital laboratories and private laboratories. The diagrams produced are extracted from the different phases of TOGAF (namely *A – Architecture Vision*, *B – Business Architecture* and *C – Data Architecture*) and have been developed to illustrate a number of constructs of the five TOGAF layers, namely *Business Architecture*, *Data Architecture*, *Application Architecture*, *Technology Architecture*, and *Architecture Principles, requirements, and roadmap*. The example merely illustrates the use of TOGAF 9 constructs (which are very generic) in a context of ISSRM.

First, to give a brief overview of the various key players, the Stakeholder Map matrix (Table 1) from phase *A – Architecture vision*, lists different stakeholders. For each stakeholder identified, key concerns, class (i.e., combination of their willingness to contribute and their power of decision) and the catalogues, matrices and diagrams they have to validate, are defined. Only an extract of the Stakeholder Map matrix established for the sector is reported in Table 1.

**Table 1: Stakeholder Map matrix**

| Stakeholder | Key concerns | Class | Catalogues, Matrices and Diagrams |
|---|---|---|---|
| **Laboratory management** | - Ensuring the competitiveness of the laboratory by ensuring high-quality services provision and developing innovative products.<br>- Meeting the needs and expectations of patients both in terms of quality of service and delays | Key Player | - Business Interaction matrix<br>- Business Footprint diagram<br>- Goal/Objective/Service diagram<br>- Organization<br>- Decomposition diagram<br>- ... |
| **Privacy Regulator** | - Ensuring that compliance with the laws and regulations in the area of personal data protection is guaranteed. | Keep Satisfied | - Business Interaction matrix<br>- Business Footprint diagram<br>- Application Communication diagram<br>- ... |
| ... | ... | ... | ... |

Thus, in the above matrix, the stakeholder "*Laboratory management*" has the willingness not only to ensure the competiveness of the laboratory by guaranteeing high-quality services provision, but in the same time to meet the needs and expectations of patients. "*Laboratory management*" can be considered as a "*Key Player*" since it has a particular interest in the project. On the other hand, the stakeholder "*Privacy Regulator*", whose key concern is to ensure that the compliance with personal data regulation is fulfilled, can be considered as a "*Keep Satisfied*" since, even it has a strong interest in the project, it is not actively involved in.

After listing the various stakeholders, the Business Service/Information Diagram extracted from the phase *B – Business Architecture* and presented in Figure 6, shows the information needed to support one or more business services. Business services (e.g., "*Prescription validation and input*") are represented by ellipses while data entities (e.g., "*Clinical information*") are represented by rectangles. Such a diagram gives an overall view of the information consumed by or produced by a business service and may also show the source of information. These requirements will be specified in lower layers including data architecture and application architecture.

According to Figure 6, the business service "*Prescription validation and input*" requires the "*Patient identity*", a "*Medical analysis prescription*", and "*Clinical information*" as information. The service mainly produces a "*Traceability sheet*" that serves as an input for the business services "*Analysis procedure selection*", "Specimen analysis", and "*Analysis procedure verification and validation*".

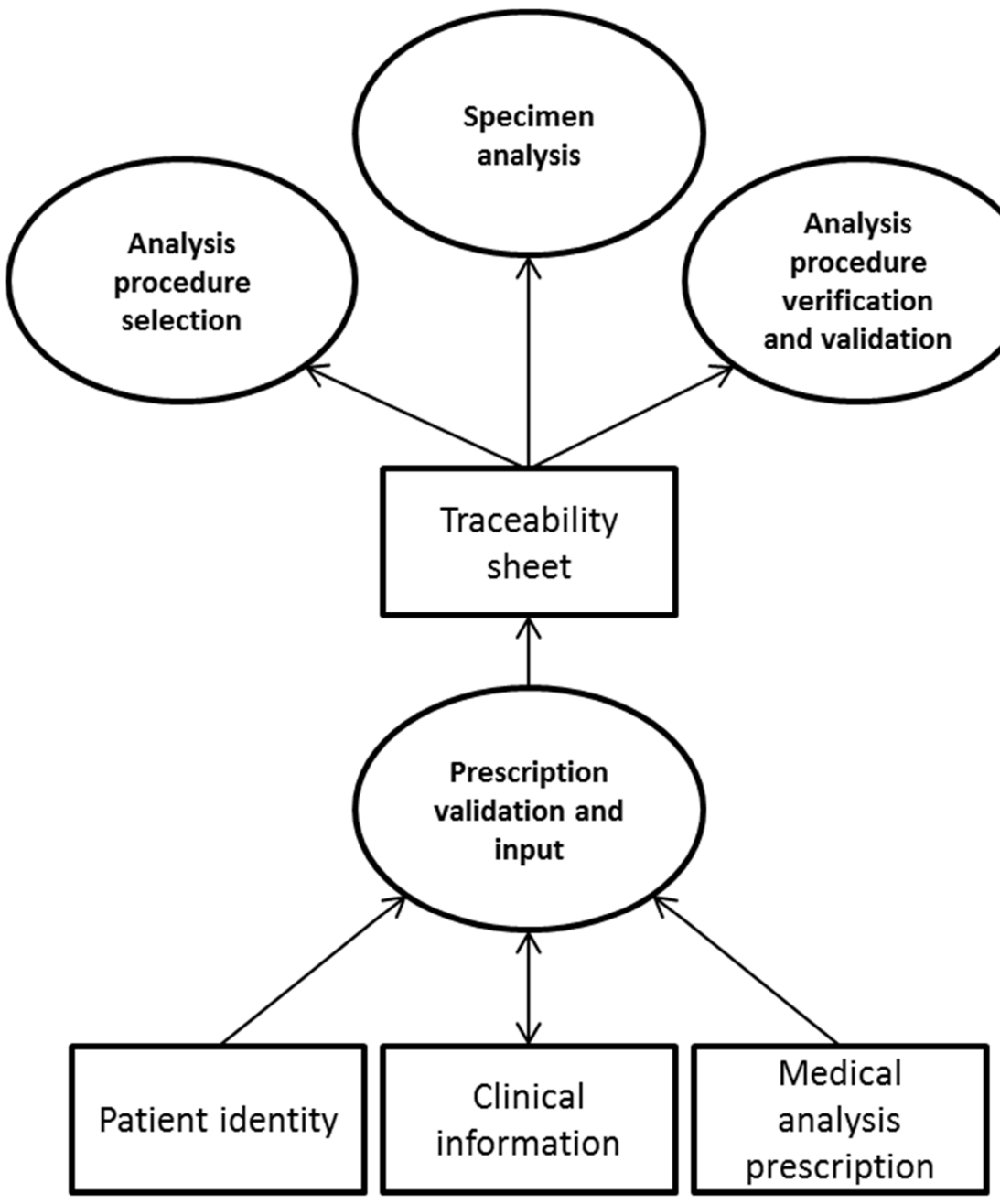

**Figure 6: Business Service/Information Diagram**

In connection with the stakeholders' key concerns (see Table 1), the Driver/Goal/Requirement catalogue (Table 2) from phase *B – Business architecture* lists the laboratories' drivers, and show how they can be met in practical terms through goals, requirements, and measures. A focus is done on the Driver "*Confidentiality*". This need for confidentiality can be expressed as a Goal: "*Confidentiality of personal information*". In order to satisfy such a goal, specific requirements may emerge for example "*Access control on biomedical analysis prescription*". Last but not least, Measures, defined as an indicator or factor that is linked to a goal or objective [6], can be associated (e.g., "*Risk of disclosure of personal data due to lack of employee's awareness*"). Indeed such an assessment may reveal weaknesses or threats that need to be addressed in order to be aligned with the related Driver "*Confidentiality*".

**Table 2: Driver/Goal/Requirement catalogue**

| Driver | Goal | Requirement | Measure |
|---|---|---|---|
| Confidentiality | Confidentiality of personal information | Access control on biomedical analysis prescription | Risk of disclosure of personal data due to lack of employee's awareness |
| ... | ... | ... | ... |

The Data Entity/Business Function Matrix extracted from the phase *C – Data Architecture* presented in Table 3 depicts the relationship between Data entities (e.g., "*Clinical information*") and Business functions (e.g., "*Business pre-analysis*") within a laboratory. Relationships are qualified using '*i*' whenever the data entity is consumed by the business function and is considered as an input, '*o*' whenever the data entity is produced by the business function and is considered as an output, or '*i/o*' when the data entity is both consumed and produced by the business function.

**Table 3: Data Entity / Business Function Matrix**

| Business Functions | Data Entities | | | | | | |
|---|---|---|---|---|---|---|---|
| | Medical Analysis Prescription | Clinical information | Patient identity | Traceability sheet | Blood specimen | Medical analysis results | Medical analysis report |
| Biomedical pre-analysis | i | i/o | i | o | o | - | - |
| Biomedical analysis | - | - | - | i | i | o | - |
| Biomedical post-analysis | - | - | - | - | - | i | o |
| Medical specimen management | - | - | - | - | i | - | - |

According to Table 3, the Business function "*Biomedical pre-analysis*" (supported among others by the Business service "*Prescription validation and input*" detailed in Figure 6) uses the following Data entities: "*Medical Analysis Prescription*", "*Patient identity*" and "*Clinical information*". It may update the "*Clinical information*", and produces a "*Traceability sheet*" as well as a "*Blood specimen*".

The more technical layers, exhibiting both application (*Application architecture*) and technology (*Technology architecture*) parts of the architecture have been deliberately set aside of this running example, mainly for their too high level of specificity and the difficulty to represent them in a concise manner, but also because their mapping with concepts of the ISSRM domain model is more straightforward.

## Alignment tables

Based on the definitions provided by the respective specification and the definitions of the concepts of the ISSRM domain [3], [4], a focus group has performed the conceptual alignment. They have built four tables depicting the structural and semantic correspondences of the concepts defined in the hereinabove four EAM methods with those of the ISSRM domain. In other words, these tables show the capabilities of each EAM method to represent the ISSRM concepts. It shall be read: "*Concept*" is a "*Semantic mapping type*" of "*ISSRM domain model concept*". For example: *Product* is a *specialisation* of *Business asset*. When applicable, each mapping is illustrated with the running example (when the concept is not exploited in the running example, a "n/a" label is put in the corresponding cell of the "Running example" column).

## ArchiMate

| ArchiMate 2.1 | Definition | ISSRM domain model | Semantic mapping types [10] | Running example |
|---|---|---|---|---|
| ***Business Layer Metamodel*** | | | | |
| **Value** | The relative worth, utility, or importance of a business service or product. | Business asset::value | equivalence | "Home care" |
| **Product** | A coherent collection of services, accompanied by a contract/set of agreements, which is offered as a whole to (internal or external) customers. | Business asset | specialisation | "Home Blood Analysis" |
| **Contract** | A formal or informal specification of an agreement that specifies the rights and obligations associated with a product. | Business asset | specialisation | "Contract" |
| **Business Object** | A passive element that has relevance from a business perspective. | Asset | specialisation | "Biomedical Analysis Prescription" |
| **Meaning** | The knowledge or expertise present in a business object or its representation, given a particular context. | Business asset | specialisation | "Prescribed Analyses" |
| **Representation** | A perceptible form of the information carried by a business object. | IS asset | specialisation | "Biomedical Paper Prescription" |
| **Business Service** | A service that fulfills a business need for a customer (internal or external to the organization). | Business asset | specialisation | "Home Blood Taking" |
| **Business Process** | A behavior element that groups behavior based on an ordering of activities. It is intended to produce a defined set of products or business services). | Business asset | specialisation | "Take Blood Home" |
| **Function** | A behavior element that groups behavior based on a chosen set of criteria (typically required business resources and/or competences). | Business asset | specialisation | “BioMedical PreAnalysis” |
| **Interaction** | A behavior element that describes the behavior of business collaboration. | Business asset | specialisation | n/a |
| **Business Event** | Something that happens (internally or externally) and influences behavior. | N/A | specialisation | n/a |
| **Business Interface** | A point of access where a business service is made available to the environment. | IS asset | specialisation | “Home Analysis Call Center” |
| **Business Role** | The responsibility for performing specific behavior, to which an actor can be assigned. | Business asset | specialisation | “Nurse” |
| **Business Collaboration** | An aggregate of two or more business roles that work together to perform collective behavior. | Business asset | specialisation | n/a |
| **Location** | A conceptual point or extent in space. | IS asset | specialisation | “Patient Home” |
| **Business Actor** | An organizational entity that is capable of performing behavior. | IS asset | specialisation | “An independent nurse” |
| ***Application layer*** | | | | |
| **Data Object** | A passive element suitable for automated processing. | IS asset | specialisation | “Biomedical Prescription Data” |
| **Application Service** | A service that exposes automated behavior. | IS asset | specialisation | “Prescription Input” |
| **Application Function/ Interaction** | A behavior element that groups automated behavior that can be performed by an application component.<br><br>A behavior element that describes the behavior of an application collaboration. | IS asset | specialisation | “Prescription Management” |
| **Application Interface** | A point of access where an application service is made available to a user or another application component. | IS asset | specialisation | n/a |
| **Application Component** | A modular, deployable, and replaceable part of a software system that encapsulates its behavior and data and exposes these through a set of interfaces. | IS asset | specialisation | "Mobile Prescription Management" |
| **Application Collaboration** | An aggregate of two or more application components that work together to perform collective behavior. | IS asset | specialisation | "Prescription Mobile App" |
| ***Technology layer*** | | | | |
| **Artifact** | A physical piece of data that is used or produced in a software development process, or by deployment and operation of a system. | IS asset | specialisation | "Prescription Mobile Application" |
| **Infrastructure Service** | An externally visible unit of functionality, provided by one or more nodes, exposed through well-defined interfaces, and meaningful to the environment. | IS asset | specialisation | n/a |
| **Infrastructure Function** | A behavior element that groups infrastructural behavior that can be performed by a node. | IS asset | specialisation | n/a |
| **Infrastructure Interface** | A point of access where infrastructure services offered by a node can be accessed by other nodes and application components. | IS asset | specialisation | n/a |
| **Node** | A computational resource upon which artifacts may be stored or deployed for execution. | IS asset | specialisation | "Mobile Node" |
| **System Software** | A software environment for specific types of components and objects that are deployed on it in the form of artifacts. | IS asset | specialisation | "Mobile OS" |
| **Device** | A hardware resource upon which artifacts may be stored or deployed for | IS asset | specialisation | "Tablet" |
| | execution. | | | |
| **Communication Path** | A link between two or more nodes, through which these nodes can exchange data. | IS asset | specialisation | n/a |
| **Network** | A communication medium between two or more devices. | IS asset | specialisation | "WiFi" |
| ***Motivation Extension*** | | | | |
| **Structure Element** | An abstract entity that is useful for classifying entity types, but not instantiated. It does not have a direct mapping to the TOGAF standard. | N/A because not instantiated | association | |
| **Stakeholder** | The role of an individual, team, or organization (or classes thereof) that represents their interests in, or concerns relative to, the outcome of the architecture. | Asset | | "Privacy Regulator" |
| **Motivational Element** | An abstract entity that is useful for classifying entity types, but not instantiated. It does not have a direct mapping to the TOGAF standard. | N/A because not instantiated | generalisation | |
| **Driver** | Something that creates, motivates, and fuels the change in an organization. | Security criterion | generalisation | "Confidentiality" |
| **Assessment** | The outcome of some analysis of some driver. | Risk | generalisation | "Risk of disclosure of personal data due to lack of employee's awareness" |
| **Goal** | An end state that a stakeholder intends to achieve. | Security objective | association | "Confidentiality of Personal Information" |
| **Principle** | A normative property of all systems in a given context, or the way in which they are realized. | Asset | generalisation | "European Personal Data Privacy Directive" |
| **Requirement** | A statement of need that must be realized by a system. | Security requirement | generalisation | "Access control on biomedical |
| | | | | analysis prescription” |
| **Constraint** | A restriction on the way in which a system is realized. | Security requirement | generalisation | n/a |

## TOGAF

| TOGAF 9.1 | Definition | ISSRM domain model | Semantic mapping types [10] | Running example |
|---|---|---|---|---|
| ***Business Architecture*** | | | | |
| **Organization Unit** | A self-contained unit of resources with goals, objectives, and measures. Organization units may include external parties and business partner organizations. | IS asset<br>Asset | specialisation<br>association | “Biomedical laboratory” |
| **Actor** | A person, organization, or system that has a role that initiates or interacts with activities; for example, a sales representative who travels to visit customers. Actors may be internal or external to an organization. In the automotive industry, an original equipment manufacturer would be considered an actor by an automotive dealership that interacts with its supply chain activities. | IS asset | specialisation | N/A |
| **Function** | Delivers business capabilities closely aligned to an organization, but not necessarily explicitly governed by the organization. Also referred to as “*Business function*”. | Business asset | specialisation | “Biomedical pre-analysis” |
| **Role** | The usual or expected function of an Actor, or the part somebody or something plays in a particular action or event. An Actor may have a number of roles. | Business asset<br>Asset | specialisation<br>association | N/A |
| **Process** | A process represents flow of control between or within functions and/or services (depends on the granularity of definition).<br>Processes represent a sequence of activities that together achieve a specified outcome, can be decomposed into sub-processes, and can show operation of a function or service (at next level of detail). Processes | Business asset | specialisation | N/A |
| | may also be used to link or compose organizations, functions, services, and processes. | | | |
| **Business Service** | Supports business capabilities through an explicitly defined interface and is explicitly governed by an organization. | Business asset | specialisation | “Prescription validation and input” |
| **Driver** | An external or internal condition that motivates the organization to define its goals. An example of an external driver is a change in regulation or compliance rules which, for example, require changes to the way an organization operates; i.e., Sarbanes-Oxley in the US. | Security criterion | generalisation | “Confidentiality” |
| **Goal** | A high-level statement of intent or direction for an organization. Typically used to measure success of an organization | Security objective | generalisation | “Confidentiality of personal information” |
| **Objective** | A time-bound milestone for an organization used to demonstrate progress towards a goal; for example, “*Increase Capacity Utilization by 30% by the end of 2009 to support the planned increase in market share*”. | Security objective | generalisation | N/A |
| **Measure** | An indicator or factor that can be tracked, usually on an ongoing basis, to determine success or alignment with objectives and goals. | Risk | generalisation | “Risk of disclosure of personal data due to lack of employee’s awareness” |
| **Location** | A place where business activity takes place and can be hierarchically decomposed. | IS asset | specialisation | N/A |
| **Event** | An organizational state change that triggers processing events may originate from inside or outside the organization and may be resolved inside or outside the organization. | N/A | N/A | N/A |
| **Product** | Output generated by the business. The business product of the execution of a process. | Business asset | specialisation | N/A |
| **Control** | A decision-making step with accompanying decision logic used to determine execution approach for a process or to ensure that a process complies with governance criteria. For example, a sign-off control on the purchase request processing process that checks whether the total value of the request is within the sign-off limits of the requester, or whether it needs escalating to higher authority. | Business asset | specialisation | N/A |
| **Service quality** | A present configuration of non-functional attributes that may be assigned to a service or service contract. | Business asset | specialisation | N/A |
| **Contract** | An agreement between a service consumer and a service provider that establishes functional and non-functional parameters for interaction | Business asset | specialisation | N/A |
| ***Data Architecture*** | | | | |
| **Data Entity** | An encapsulation of data that is recognized by a business domain expert as a thing. Logical data entities can be tied to applications, repositories, and services and may be structured according to implementation considerations. | IS asset | specialisation | “Clinical information” |
| **Physical Data Component** | A boundary zone that encapsulates related data entities to form a physical location to be held. For example, a purchase order business object, comprising purchase order header and item business object nodes. | IS asset | specialisation | N/A |
| **Logical Data Component** | A boundary zone that encapsulates related data entities to form a logical location to be held; for example, external procurement information. | IS asset | specialisation | N/A |
| ***Application Architecture*** | | | | |
| **Information** | The automated elements of a business service. An information system | IS asset | specialisation | N/A |
| **System Service** | service may deliver or support par t or all of one or more business services | | | |
| **Logical Application Component** | An encapsulation of application functionality that is independent of a particular implementation. For example, the classification of all purchase request processing applications implemented in an enterprise. | IS asset | specialisation | N/A |
| **Physical Application Component** | An application, application module, application service, or other deployable component of functionality. For example, a configured and deployed instance of a Commercial Off-The-Shelf (COTS) Enterprise Resource Planning (ERP) supply-chain management application. | IS asset | specialisation | N/A |
| ***Technology Architecture*** | | | | |
| **Logical Technology Component** | An encapsulation of technology infrastructure that is independent of a particular product. A class of technology product; for example, supply chain management software as part of an Enterprise Resource Planning (ERP) suite, or a Commercial Off-The-Shelf (COTS) purchase request processing enterprise service. | IS asset | specialisation | N/A |
| **Platform Service** | A technical capability required to provide enabling infrastructure that supports the delivery of applications | IS asset | specialisation | N/A |
| **Physical Technology component** | Physical Technology Component A specific technology infrastructure product or technology infrastructure product instance. For example, a particular product version of a Commercial Off-The-Shelf (COTS) solution, or a specific brand and version of server. | IS asset | specialisation | N/A |
| **Technology Component** | An encapsulation of technology infrastructure that represents a class of technology product or specific technology product. | N/A | N/A | N/A |
| **Architecture Principles, Requirements, and Roadmap** | | | | |
| **Principle** | A qualitative statement of intent that should be met by the architecture. Has at least a supporting rationale and a measure of importance. | Asset | association | N/A |
| **Constraint** | An external factor that prevents an organization from pursuing particular approaches to meet its goals. For example, customer data is not harmonized within the organization, regionally or nationally, constraining the organization's ability to offer effective customer service. | Asset | association | N/A |
| **Assumption** | A statement of probable fact that has not been fully validated at this stage, due to external constraints. For example, it may be assumed that an existing application will support a certain set of functional requirements, although those requirements may not yet have been individually validated. | Asset | association | N/A |
| **Requirement** | A quantitative statement of business need that must be met by a particular architecture or work package. | Security requirement | generalisation | "Access control on biomedical analysis prescription" |
| **Gap** | A statement of difference between two states. Used in the context of gap analysis, where the difference between the Baseline and Target Architecture is identified. | N/A | N/A | N/A |
| **Work Package** | A set of actions identified to achieve one or more objectives for the business. A work package can be a part of a project, a complete project, or a program. | N/A | N/A | N/A |
| **Capability** | An ability that an organization, person, or system possesses. Capabilities are typically expressed in general and high level terms and typically require a combination of organization, people, processes, and technology to achieve.<br>A business-focused outcome that is delivered by the completion of one or more work packages. Using a capability-based planning approach, | Business asset | specialisation | N/A |
| | change activities can be sequenced and grouped in order to provide continuous and incremental business value. | | | |
| **Other** | | | | |
| **Service** | An element of behavior that provides specific functionality in response to requests from actors or other services. A service delivers or supports business capabilities, has an explicitly defined interface, and is explicitly governed. Services are defined for business, information systems, and platforms. | N/A | N/A | N/A |

### DoDAF

| DoDAF 2.02 | Definition | ISSRM domain model | Semantic mapping types [10] |
|---|---|---|---|
| **Activity** | Work, not specific to a single organization, weapon system or individual that transforms inputs (Resources) into outputs (Resources) or changes their state | Business asset | specialisation |
| **Resource** | Data, Information, Performers, Materiel, or Personnel Types that are produced or consumed. | Asset | specialisation |
| **Materiel** | Equipment, apparatus or supplies that are of interest, without distinction as to its application for administrative or combat purposes | IS asset | specialisation |
| **Information** | Information is the state of a something of interest that is materialized -- in any medium or form -- and communicated or received | Business Asset | specialisation |
| **Data** | Representation of information in a formalized manner suitable for communication, interpretation, or processing by humans or by automatic means. Examples could be whole models, packages, entities, attributes, classes, domain values, enumeration values, records, tables, rows, columns, and fields | IS asset | specialisation |
| **Architecture Description** | Information describing an architecture such as an OV-5 Activity Model document. | N/A | N/A |
| **Performer** | Any entity - human, automated, or any aggregation of human and/or automated - that performs an activity and provides a capability | IS asset | specialisation |
| **Organization** | A specific real-world assemblage of people and other resources organized for an on-going purpose | IS asset | specialisation |
| **System** | A functionally, physically, and/or behaviorally related group of regularly interacting or | IS asset | specialisation |
| | interdependent elements. | | |
| **Person(nel) Role /Person Type**[2] | A category of person roles defined by the role or roles they share that are relevant to an architecture. Includes assigned materiel. | IS asset | specialisation |
| **Service** | A mechanism to enable access to a set of one or more capabilities , where the access is provided using a prescribed interface and is exercised consistent with constraints and policies as specified by the service description. The mechanism is a Performer. The "capabilities" accessed are Resources -- Information, Data, Materiel, Performers, and Geo-political Extents. | IS asset | specialisation |
| **Capability** | The ability to achieve a Desired Effect under specified [performance] standards and conditions through combinations of ways and means [activities and resources] to perform a set of activities. | IS asset + Business asset | equivalence |
| **Condition** | The state of an environment or situation in which a Performer performs or is disposed to perform | Asset | association |
| **Desired Effect** | A desired state of a Resource | Security objective | generalisation |
| **Measure** | The magnitude of some attribute of an individual | Attributes of the concepts | equivalence |
| **Measure Type** | A category of Measures | N/A | N/A |
| **Location** | A point or extent in space that may be referred to physically or logically | IS asset | specialisation |
| **Guidance** | An authoritative statement intended to lead or steer the execution of actions | Asset | association |
| **Rule** | A principle or condition that governs behavior; a prescribed guide for conduct or action | Asset | association |
| **Agreement** | A consent among parties regarding the terms and conditions of activities that said parties | Asset | association |

[2] Named PersonRole in diagram – Person Type/Personnel Type in the definitions

| | | | |
|---|---|---|---|
| | participate in. | | |
| **Standard** | A formal agreement documenting generally accepted specifications or criteria for products, processes, procedures, policies, systems, and/or personnel | Asset | association |
| **Project** | A temporary endeavor undertaken to create Resources or Desired Effects | N/A | N/A |
| **Vision** | An end that describes the future state of the enterprise, without regard to how it is to be achieved; a mental image of what the future will or could be like | N/A | N/A |
| **Skill** | The ability, coming from one's knowledge, practice, aptitude, etc., to do something well (potentially related terms: Training, Knowledge, Ability – is part of the PersonRole) | Business asset | specialisation |
| **GeoPolitical Extent** | A geospatial extent whose boundaries are by declaration or agreement by political parties. | IS asset | specialisation |

### IAF

| IAF | Definition | ISSRM domain model | Semantic mapping types [10] |
|---|---|---|---|
| ***Business Architecture*** | | | |
| **Business object** | A business object is a physical resource used by the business that is significant to the architecture. Typical business objects are *containers & trucks* (transport industry), *oil & steel* (manufacturing industry), and *contracts & money* (financial industry) | IS asset (if carries information)<br><br>N/A (if does not carry information) | specialisation |
| **Object contracts** | Object contracts describe how business services use business objects, e.g. *reading, writing or transforming an insurance proposal*. | Business asset<br><br>*Attribute*::Business Asset | specialisation |
| **Business event** | In computing an event is an action that is usually initiated outside of a system and has to be handled by the system. Business events therefore are actions that the business and its supporting IT must react on.<br>Examples of business events are "*order placement of an article by a customer*", "*request of a quote by that same customer"*, or "*the receipt of a payment from a customer via the bank*". | N/A | |
| **Business activity** | A business activity is a business task or group of business tasks that are undertaken by the business to achieve a well defined goal. Business activities are a description of "*WHAT*" the business does in order to meet its goals. They are implementation independent (i.e. independent of any organizational structure or process) and have clearly defined objectives in transforming an initial state to another state. A typical business service in the oil industry would be "*interpreting geographic data to find possible oil reserves"*. | Business asset | specialisation |
| **Business goal** | Business goals describe what the business needs to achieve in order to fulfill its business objectives. A business goal is an implementation independent, fundamental, and unique | Business asset | specialisation |
| | contribution to the business mission. The business goal is the "*WHY*" objective for any business activity.<br>Business goals provide a reference baseline for comparing current state and future desired state. They support the definition of results related targets for the organization. The goal of "*interpreting geographic data to find possible oil reserves*" is probably "*finding new oil reserves*". | | |
| **Business role** | A business role performs a business activity. Roles may also have accountabilities for goals (although there will be corresponding governance activities for those goals). Roles should not be associated with people or systems as people have multiple roles. Roles are independent of implementation but are still needed to support the activities. Roles relate to specific activities and support the same business goal as the activity. | Security objective | specialisation |
| **Business service** | A Business service characterizes a unique "element of business behaviour" in terms of a business activity, undertaken by a specific role that together support a specific business goal. Business objects are used by business services. The way they are used is documented through object contracts. Business events trigger business services, which in turn can trigger other business services to provide the requested result. | Business asset | specialisation |
| **Business domain** | Value chains (or parts of them), parts of an organization, and other subject areas of a business can be positioned as business domains. Usually they consist of a collection of business services contributing to a joint, certain higher business goal. | Business asset | specialisation |
| **Business Service Collaboration Contract** | Business services interact with each other. These interactions and their nature are described in business service collaboration contracts. | Business asset<br><br>*Attribute*::Business Asset | specialisation |
| **Physical business component** | Physical architecture is all about mapping logical components to real life, tangible physical components. In this case we are talking about business components, so what we need to do is allocate them to real life physical business elements that will be responsible for delivering the services that are contained in the components. | IS asset | specialisation |
| **Business Standards, Rules and Guidelines** | Physical business components and their collaboration are just one part of the physical business architecture. Just as important are the business standard, rules, and guidelines (SRGs). In effect they are a list of the topics that you will use to ensure a business architecture's implementation is done the way you want it to be done. In other words they are the criteria you will use to validate business architecture's implementation. | Security criterion | generalisation |
| **Business tasks (specifications)** | Business tasks are job, role, and task descriptions, based on the identified roles and actors. They are usually required (even often requested by the human resources department) to find and align resources which will perform these business tasks. | Business assets | specialisation |
| **Business Migration Specifications/Implementation Guidelines** | There might be circumstances in which you do not want to document the way the business should migrate in terms of standards, rules, and guidelines. The most common reason is the existence of a formal process for approving SRGs within an organization, which might take a lot of time and thus delay the architecture's implementation. In that case you can document business migration specifications and pass them on to the implementers. | N/A (related to project mgt aspects of EA) | N/A |
| ***Information architecture*** | | | |
| **Information object** | An information object is the subject of communication for business services. The information object describes the information used or communicated by business information services. An information object is a source of information. It is not a description of data but rather indicates where data is used. An information object is independent of the media it is presented on. Information objects are characterized by statements that have the general form of:<br>A "*Blah*" is a "*blur*" that "*bleeps*", for example: STATEMENT: An ORDER is the request of a CUSTOMER to supply an ARTICLE. | Business asset | specialisation |
| **Business information service** | A business information service is a construct of a business service and the information objects it uses. A business service uses information objects as input (get), or changes (transform) or creates (write) them. Thus a business information service is a business service for which the relationship to information objects has been defined. […] | Business asset | specialisation |
| | A business information service changes the perspective on a business service. We are now looking especially at the information aspect of a business service. | | |
| **Business information service collaboration contract** | Normally this is the same as the business service collaboration contract [..]. If it is relevant, you can add information processing oriented attributes to the contract. For example you can indicate which objects are passed between the business information services, or indicate which messages are part of the communication. Adding this type of information to the contract is normally only done in solution level architectures. In enterprise/domain level architectures it is done in less. | Business asset<br>*Attribute*::Business Asset | specialisation |
| **Information domain** | Information domains are used to communicate information objects to any stakeholder or groups of stakeholders. The domains are groupings of information objects according to some criteria. | Business asset | specialisation |
| **Logical information component** | Business information services use information objects in different ways (get, transform, write). While constructing logical information components you look at the interdependencies of information objects from a business information service point-of-view. In other words you find out which business information services are necessary to get/transform/write information objects in order to get/transform/write other information objects. Clustering these information objects around their relationship to similar business information services groups will lead to groups of information objects, called logical information components. | Business asset | specialisation |
| **Logical business information component** | Logical business information components are created to check if the logical business components we have defined do not violate some obvious information processing rules. The information processing rules we talk about here are: (1) You cannot use something that has not been created yet and (2) If somebody else changes something you need to use, they better change it before you want to use it. | Business asset | specialisation |
| **Logical business information** | Normally this is the same as the business information service collaboration contracts [...]. You can investigate if collaboration contracts can be merged. | Business asset<br>*Attribute*::Business Asset | specialisation |
| **component collaboration contract** | | | |
| **Physical Information Component** | Here we allocate the logical information components to real life physical entities that will be responsible for managing and storing the objects within the components. This could be the allocation of information components to physical locations or organizational units. | IS asset | specialisation |
| **Information migration specifications** | Just as in business architecture, there will be circumstances in which you want to pass on instructions regarding the information architecture to the implementers, without turning them into SRGs. They can address similar topics as will be mentioned within the SRGs, but will be less formal. Typical examples of migration specifications are related to the order in which different information objects need to be migrated to ensure data integrity and specifications regarding information conversion. | N/A (related to project mgt aspects of EA) | N/A |
| **Information Standards, Rules, and Guidelines** | Information Standards, Rules, and Guidelines (SRGs) document what the implementation of the information architecture needs to adhere to. Typical examples are:<br>• Policies for email, communication with external organizations etc.;<br>• Policies for backup, integrity, availability, confidentiality;<br>• Legislative rules for archive, access, audit;<br>• Legislative rules regarding privacy, usage of specific forms etc.;<br>• Information standards, corporate and/or industry specific. | Security criterion | generalisation |
| ***Information system architecture*** | | | |
| **Information system service** | We analyze the Business information services (within the components) and define the corresponding IS services. This sounds simple, but there is a catch to it. A simple business Information service like "*accept claim"* can lead to a lot of decision making. Are we going to accept claims via e-mail? Are we going to accept them via paper mail? If we accept them via paper mail does it have to be on a typed form so we can scan and OCR the claim, or do we also accept hand-written claims? Do we want to accept claims through the phone? | IS asset | specialisation |
| | All this decision making requires collaboration between the business, information and IS architect. | | |
| **Information system domain** | Information system domains are used to communicate the IS services. […]<br>If IS specific domains need to be defined, the following criteria can be used as a basis:<br>• Type of technology, e.g. all internet IS services are shown together;<br>• Complexity of the IS services e.g. easy to create, hard to create;<br>• Level of automation e.g. the IS service fully automates the business information service, or partly or marginally. | IS asset | specialisation |
| **Information system service collaboration contract** | Collaboration contracts between IS services are derived from the collaboration contracts between the business information services. Attributes like growth, service windows and quality of information can be copied from the business information service collaboration contract. Other attributes like response time, throughput and peak characteristics will have to be derived. | IS asset<br><br>*Attribute*::IS Asset | specialisation |
| **Logical information system component** | A Logical Information System Component is the basic element of an "ideal" or "to be" application structure created by the grouping of one or more IS services. | IS asset | specialisation |
| **Logical Information System Component Collaboration Contract** | The LISC collaboration contracts are derived from the IS Service collaboration contracts. When needed you can add attributes to the collaboration contract. | IS asset<br><br>*Attribute*::IS Asset | specialisation |
| **Physical Information System Component** | Once you have allocated the logical IS components to physical, real life things you can buy or build, you have created the physical IS components. In our case we were able find a package based solution for all of our Logical IS components. | IS asset | specialisation |
| **Physical Information System Component Collaboration Contract** | The interaction between two Physical Information System Components is documented in the Physical Information System Component Collaboration Contract. Attributes are derived from the logical IS component collaboration contract. If needed you can add any attributes you require, just as described in the logical level. | IS asset<br><br>*Attribute*::IS Asset | specialisation |
| **Information System Standards, Rules and Guidelines** | This key artifact in the IS architecture is the "law" for anybody that will be using or implementing the architecture. It can – and often will – refer to generic SRGs that have been created for IS architecture. On top of that you should define which SRGs are specific to the implementation and usage of this IS architecture. | Security requirement | generalisation |
| ***Technology infrastructure architecture*** | | | |
| **Technology infrastructure service** | Technology infrastructure services are commonly derived from IS services or logical IS components. This does not mean that you always have to create an IS architecture before you can create the TI architecture. You can derive TI services from Business services if you need to. You will only have to assume more about the IS support that will be required. | IS asset | specialisation |
| **Technology infrastructure service collaboration contract** | TI service collaboration contract attributes are derived from the IS service collaboration contracts. | IS asset | specialisation |
| **Technology infrastructure domain** | Just like the other domains, the TI domains are used to communicate the TI services that have been defined. | IS asset | specialisation |
| **Logical technology infrastructure component** | Logical technology infrastructure components are created using the same technique as all other component types. Services are grouped into components based on criteria that have been derived from the architecture principles. | IS asset | specialisation |
| **Logical technology infrastructure component collaboration contract** | Most of the attributes that are defined in the standard collaboration contract are also relevant for technology infrastructure collaboration contracts. Some might be less relevant and can be removed. | IS asset<br>*Attribute*::IS Asset | specialisation |
| **Physical technology infrastructure component** | Products or technologies chosen to realize a Logical Technology Infrastructure Components are called Physical Technology Infrastructure Components. Possible examples could be HP UX, Oracle, Cisco's network appliances or Microsoft's Office Suite. Their specifications can be documented using the component attributes. Very often they are visualized using pictures from real life. | IS asset | specialisation |
| **Physical technology infrastructure component collaboration contract** | These are derived from the logical TI component collaboration contracts. They commonly have the same attributes. | IS asset<br><br>*Attribute*::IS Asset | specialisation |
| **Technology Infrastructure Standards, Rules and Guidelines** | Just as in the IS architecture, this is a key artifact in the TI architecture. It is defined for the same purpose, it is the "law" for anybody that will be using or implementing the architecture. | Security requirement | generalisation |
| **Technology Infrastructure Migration Specifications** | Just as in all the other aspect areas, there will be circumstances in which you want to pass on instructions regarding the TI architecture to the implementers, without turning them into SRGs. They can address the same topics as will be mentioned within the SRGs, but will be less formal. Typical examples of migration specifications are related to (1) the order in which different technology infrastructure components objects need to be migrated to ensure overall transaction integrity, (2) the rollback possibilities and implications, and (3) the deployment steps. | N/A (related to project mgt aspects of EA) | N/A |
| ***Quality aspects of architecture*** | | | |
| **Logical components** | Some logical components have their main reason of existence justified in governance or security reasons. This holds true for all IAF aspect areas. This could take the form of an actor such as "security officer", "compliance officer" or the business component "IT operations". On the information side it could be the CMDB (Configuration Management | Control | equivalence |
| | DataBase) or the set of compliance rules.<br>In other cases these type of components have over time become elements of the technology infrastructure. Examples of these common security technology components are: directory server, virus checker, access manager, firewall, etc.. | | |
| **Control** | A control is a means of managing the risk of a required level of quality being compromised. […] The term “control” is also used as a synonym for a safeguard, security measure, countermeasure or mitigation. […] Controls can be found in all aspect areas of IAF and even outside IAF. Some controls are applicable to the business aspect area, as can be seen from examples such as segregation of duties, staff screening, authorization, and monitoring, logging and auditing business services. For the information aspect an organization might consider classification of information, encrypting information or keeping copies of information. Technology – wise it could be automated intrusion detection, redundant equipment, smartcards to store authorizations and identities, and so on. True physical controls should also be considered. | Control | equivalence |
| **Physical components** | Physical components are allocated to the logical components. If we find that the physical components cannot completely fulfill the specifications of the logical components, we have encountered a gap. | Control | equivalence |
| **Quality Standards, Rules and Guidelines** | The architect specifies additional standards, rules and guidelines to ensure that the required level of quality is guaranteed. These SRGs can take the form of specific (quality, security, governance ...) standards, rules or guidelines, or they can be added to the SRGs specified for the aspect areas. | Security requirement | generalisation |
| **Service Level Agreement (SLA)** | A Service Level Agreement (SLA) is as a physical contract between two parties, in which one party agrees to deliver services to another with a guaranteed level of service. The quality of the services needed are defined in the logical contracts. The actual level of service that is agreed upon in the SLA might differ for reasons of feasibility, cost, etc. | Control | specification |

# Bibliography and References

# About the authors

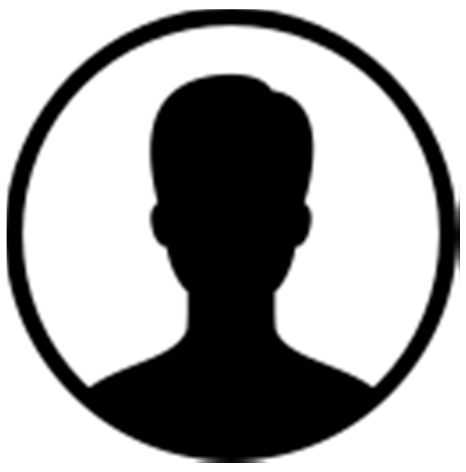

Dr. Nicolas Mayer
nicolas.mayer@list.lu

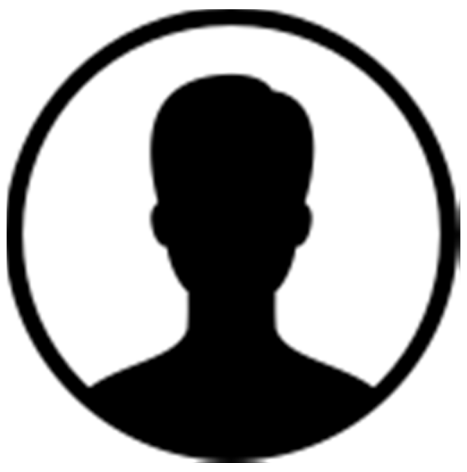

Jocelyn Aubert
jocelyn.aubert@list.lu

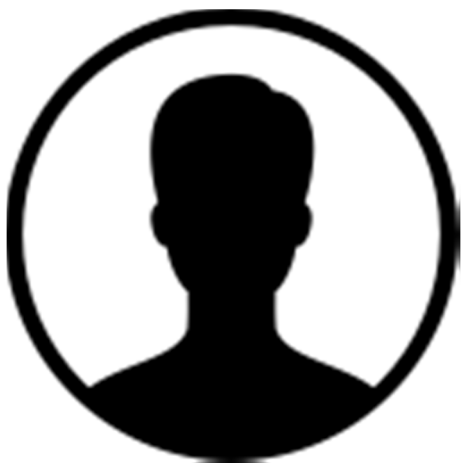

Eric Grandry
eric.grandry@list.lu

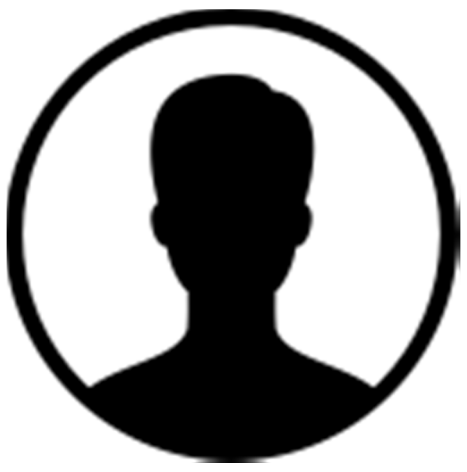

Dr. Christophe Feltus
christophe.feltus@list.lu

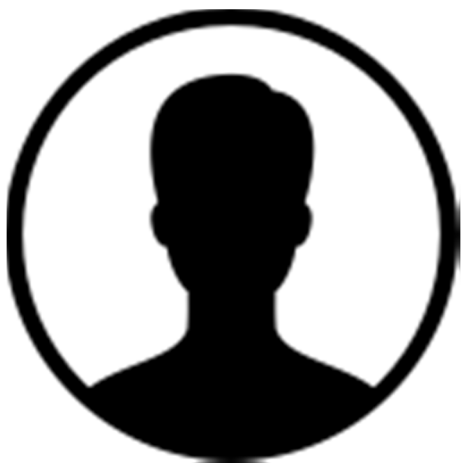

Dr. Elio Goettelmann
elio.goettelmann@list.lu